\documentclass[aps,showpacs]{revtex4}%
\usepackage{amsmath}
\usepackage{graphicx}%
\usepackage{amsfonts}%
\usepackage{amssymb}
\providecommand{\U}[1]{\protect\rule{.1in}{.1in}}
\providecommand{\U}[1]{\protect\rule{.1in}{.1in}}
\providecommand{\U}[1]{\protect\rule{.1in}{.1in}}
\providecommand{\U}[1]{\protect\rule{.1in}{.1in}}
\providecommand{\U}[1]{\protect\rule{.1in}{.1in}}
\begin{document}
\title{ Dimensional reduction as a method to obtain dual theories for massive spin
two in arbitrary dimensions}
\author{A. Khoudeir$^{1}$, R. Montemayor$^{2}$ and Luis F. Urrutia$^{3}$}
\affiliation{$^{1}$ Centro de F\'\i sica Fundamental, Departamento de F\'\i sica, Facultad
de Ciencias , Universidad de Los Andes, M\'erida 5101, Venezuela}
\affiliation{$^{2}$ Instituto Balseiro and CAB, Universidad Nacional de Cuyo and CNEA, 8400
Bariloche, Argentina}
\affiliation{$^{3}$ Instituto de Ciencias Nucleares, Universidad Nacional Aut{\'o}noma de
M{\'e}xico, A. Postal 70-543, 04510 M{\'e}xico D.F.}

\begin{abstract}
Using the parent Lagrangian method together with a dimensional reduction from
$D$ to $(D-1)$ dimensions we construct dual theories for massive spin two
fields in arbitrary dimensions in terms of a mixed symmetry tensor
$T_{A[A_{1}A_{2}\dots A_{D-2}]}$. Our starting point is the well studied
massless parent action in dimension $D$. The resulting massive
Stueckelberg-like parent actions in $(D-1)$ dimensions inherits all the gauge
symmetries of the original massless action and can be gauge fixed in two
alternative ways, yielding the possibility of having either a parent action
with a symmetric or a non-symmetric Fierz-Pauli field $e_{AB}$. Even though
the dual sector in terms of the standard spin two field includes only the
symmetrical part $e_{\{AB\}}$ in both cases, these two possibilities yield
different results in terms of the alternative dual field $T_{A[A_{1}A_{2}\dots
A_{D-2}]}$. In particular, the non-symmetric case reproduces the
Freund-Curtright action as the dual to the massive spin two field action in
four dimensions.

\end{abstract}

\pacs{11.10.-z, 11.90.+t, 02.90.+p}
\maketitle


\section{Introduction}

The fact that in dimension $D>5$ the totally symmetric tensor fields are not
enough to cover all the irreducible representations of the Poincar\'{e} group
has motivated the study of fields with mixed symmetry \cite{CURT, CURT1}
belonging to \textquotedblleft exotic\textquotedblright\ representations of
the Poincar\'{e} group. Additional interest in such fields arises because it
is quite natural to expect that in the low energy limit the superstring theory
should reduce to a consistent interacting supersymmetric theory of massless
and massive higher spin fields ($s\geq2$) arising from higher dimensions. This
proliferation of \textquotedblleft exotic\textquotedblright\ mixed symmetry
fields poses the question of identifying different representations that can
describe the same spin, possible in different phases with respect to a
weak/strong coupling limit. This is precisely the subject of duality, which
has been profusely studied along the years in many different contexts
\cite{Hull, Bekaert3}. In the massless case, dual formulations of higher spin
($s\geq2$) fields in arbitrary dimensions have been derived from a first order
parent action \cite{Boulanger3} based upon the Vasiliev action
\cite{Vasiliev2}. In this case, when the original description of the gauge
fields in dimension $D$ is in terms of totally symmetric tensors, dual
theories in terms of mixed symmetry tensors corresponding to Young tableaux
having one column with ($D-3$) boxes plus ($s-1$) columns with one box have
been obtained \cite{Boulanger3}. A discussion of duality for massless spin two
in arbitrary dimensions consistent with the Vasiliev formulation
\cite{Vasiliev2} has also been presented in Ref. \cite{West}.
Furthermore,\ the method of the global symmetry extension \cite{LINDSTROM}%
 has been applied to the dualization of massless spin two fields in arbitrary
dimensions \cite{HARIKUMAR}. An extension of these results to an AdS
background was given in Ref. \cite{Matveev}).

Dual formulations for massive higher spin fields are not as well explored.
Because massive spin two fields naturally appear in brane-world models, there
is an increasing interest in the understanding of alternative descriptions of
massive gravitons in arbitrary dimensions. Of the many approaches available to
produce dual theories we work with the parent Lagrangian method. Basically, in
the case of a spin two field, this method is based on a first order action
including both the standard linear graviton field $e_{ab}$ together with the
corresponding dual field. The individual actions are recovered after
eliminating the unwanted field using its equations of motion. In this way, on
the one hand we recover the Fierz-Pauli (FP) theory and on the other the
proposed dual formulation.\ It is known that a dimensional reduction of a
massless spin two theory in $D$\ dimensions leads to a massive spin two theory
in $(D-1)$\ dimensions \cite{STRD}. Since the parent action for massless spin
two field is known in dimension $D$, we investigate the resulting parent
action in $(D-1)$ dimensions arising from a process of dimensional reduction
by compactifying one dimension in a circle. Such a reduced parent action will
describe a massive spin two field and we will derive the corresponding dual
theory from it. Even though a mass is present, the reduced parent action
inherits all the gauge symmetries of the original massless theory in $D$
dimensions, so that we end up with a Stueckelberg-like formulation. In this
way, the resulting dual actions written in terms of the propagating fields are
only obtained after following a mixture of two steps. (1) On the one hand we
need to specify the required gauge fixings that still leave the resulting
Lagrangians in the same gauge orbit, thus making them equivalent via gauge
transformations and/or field redefinitions. This means that a unique
Lagrangian is obtained after choosing a specific point in the gauge orbit. (2)
On the other hand, and following the basic idea of the parent Lagrangian
approach, we perform a series of field eliminations via their equations of
motion. It is precisely this last process that produces inequivalent final
Lagrangians that nevertheless describe the same number of degrees of freedom.
This aspect of the construction is most clearly seen when the parent
Lagrangian has no gauge freedom and each of the fields is eliminated to
produce the corresponding non-equivalent dual actions. That is to say, we can
expect that alternative field elimination among the remaining auxiliary fields
after different gauge fixing will produce non-equivalent final dual Lagrangians.

In other words, at the level of the gauge invariant theory we only know for
sure that we have $D(D-3)/2$ independent degrees of freedom, which will
reorganize themselves according to the way the gauge and field eliminations
are selected. Hence, the method is not free from ambiguities, which basically
originate from these choices. An alternative Stueckelberg-like approach has
been developed by Zinoviev \cite{ZINOVIEV1,ZINOVIEV2} and suffers from the
same type of ambiguities. There are additional ways of compactifying the extra
dimension \cite{SIEGEL}, which are not discussed in this work.

The paper is organized as follows: in Section II we set our conventions and
the strategy to carry out the dimensional reduction from $D$ to $(D-1)$
dimensions. Also we show that such reduction produces the Fierz-Pauli theory
in $(D-1)$ dimensions when starting from the corresponding massless spin two
action in $D$ dimensions. In Section III we start from the massless parent
action of Refs. \cite{Boulanger3,West} in $D$ dimensions and dimensionally
reduce it to a massive parent action in $(D-1)$ dimensions. From this massive
parent action we show in Section IV that it is possible to obtain, via
different gauge fixings and field eliminations, alternative parent actions
containing either a symmetric ($e_{\left\{  ab\right\}  }$) or a non-symmetric
($e_{ab}$) standard spin two field. In Section V we construct the
corresponding dual theories for the massive standard spin two field in
arbitrary dimensions. In $D=4$ and for $e_{\left\{  ab\right\}  }$ we recover
one of the families described in Ref. \cite{CMU2},\ while for the
non-symmetric case we recover the action proposed in Ref. \cite{CURT}. Section
VI contains some comments, which summarize the paper. Finally in the Appendix
we set the mass parameter equal to zero in the $D=4$ massive parent action,
obtained from the massless five-dimensional one, and exhibit two different
gauge fixing, which reshuffles the original five degrees of freedom into the
sum of spin two, one and zero non interacting theories. One of such gauge
choices leads to a rather unexpected Stueckelberg-like reformulation of the
massless spin one field.

\section{Dimensional reduction of a massless $s=2$\ field from $D$ to $(D-1)$
dimensions in flat space-time}

The action for a massless spin two field in a $D$ dimensional flat space time
is%
\begin{equation}
S_{0}^{D}=\int d^{D}x\left[  -\partial_{C}e_{\left\{  BA\right\}  }%
\partial^{C}e^{\left\{  BA\right\}  }+\partial_{C}e\partial^{C}e-2\partial
^{M}e\partial^{N}e_{\left\{  NM\right\}  }+2\partial^{M}e_{\left\{
MA\right\}  }\partial_{N}e^{\left\{  NA\right\}  }\right]  ,
\end{equation}
while for a massive spin two field the action is the same plus the Fierz-Pauli
mass term%
\begin{equation}
S_{\mu}^{D}=S_{0}^{D}-\mu^{2}\int d^{D}x\left(  e_{\left\{  AB\right\}
}e^{\left\{  AB\right\}  }-e^{2}\right)  .
\end{equation}
In both cases $e^{\left\{  AB\right\}  }$ is a symmetric tensor, $e^{\left\{
AB\right\}  }=e^{\left\{  BA\right\}  }$ and we are using the metric
$diag(-,+,+...+)$.

In the massless case there is a local symmetry, related to an arbitrary change
of coordinates $x^{A}\rightarrow x^{A}+\xi^{A}(x)$, given by $e^{\left\{
AB\right\}  }\rightarrow e^{\left\{  AB\right\}  }+\left(  \partial^{A}\xi
^{B}+\partial^{B}\xi^{A}\right)  $. A complete gauge fixing implies $2D$
constraints. For example, as it is usually done in $D=4$, we can fix this
symmetry such that $\partial_{A}e^{\left\{  AB\right\}  }=0$ ($D$
constraints), but still remains a symmetry corresponding to the
transformations that maintain these relations unaltered, i.e. the ones that
satisfy $\partial_{B}\partial^{B}\xi^{A}=0$. The fixing of this last symmetry
leads to $D$ additional constraints. Thus the number of degrees of freedom for
the spin two massless field is%
\begin{equation}
f_{0}^{D}=\frac{D}{2}(D+1)-2D=\frac{D}{2}(D-3).
\end{equation}

In the case of a massive field there is no gauge symmetry due to the mass
term, but its Euler-Lagrange equations yield $(D+1)$ constraints,
$\partial_{A}e^{\left\{  AB\right\}  }=0$ and $e_{A}^{\ \ A}=e=0$, so that the
number of degrees of freedom is
\begin{equation}
f_{\mu}^{D}=\frac{D}{2}(D+1)-(D+1)=\frac{1}{2}\left(  D+1\right)  \left(
D-2\right)  .
\end{equation}
Notice that the massless spin two field in $D$ dimensions has the same number
of degrees of freedom that the massive field in $D-1$ dimensions, $f_{0}%
^{D}=f_{\mu}^{D-1}$. This suggest a relation between both fields via
dimensional reduction. This point is explored in the following.

To be specific we will consider the reduction from $D$ to $\left(  D-1\right)
$ dimensions by compactifying one of the spatial coordinates, $y$, on a circle
$S^{1}$ of radius $L$ so that the remaining space continues to be flat. We
denote the indices of the $D$ dimensional tensors with capital letters
($A,B,...M=0,1,...,D-1$)\ and reserve the lower case ones
$(a,b,...m,...=0,1,2,3,..D-2)$ to the $(D-1)$ dimensional tensors. The spatial
dimension to be reduced by compactification is denoted by the index $\left(
D-1\right)  $\ so that $A=\left(  a,D-1\right)  $ and $X^{M}=\left(
x^{m},x^{\left(  D-1\right)  }=y\right)  $.\ The basic idea in the reduction
is to rewrite any $D$ dimensional action in terms of this splitting
$A=(a,D-1)$. We expand all the fields in $D$ dimensions as a Fourier series of
the form%
\begin{equation}
\Psi_{AB...}{}^{RS...}(X^{M})=\sum_{n}\Psi^{\left(  n\right) }{}_{AB...}%
{}^{RS...}(x^{m}) e^{iny/L},
\end{equation}
and we consider a mode with $n/L=\mu$. In this case the coordinate dependence
of a $D$ dimensional real tensor $\Phi_{AB...}^{\ \ \ \ \ RS...}$ is written
as%
\begin{equation}
\Phi_{AB...}^{\ \ \ \ \ \ RS...}(X^{M})=\sqrt{\frac{\mu}{4\pi}}\Phi
_{AB...}^{\;\;\;\;\;\;RS...}(x^{m})e^{i\mu y}+\sqrt{\frac{\mu}{4\pi}}%
\Phi_{AB...}^{\ast\;\;\;\;\;\;RS...}(x^{m})e^{-i\mu y},
\end{equation}
where $\mu$ has dimension of mass and will become the mass coefficient for the
four dimensional massive fields.

The tensorial transformation under $\left(  D-1\right)  $-parity
($y\rightarrow-y$) is defined by%
\begin{equation}
\Phi_{AB...}^{\;\;\;\;\;\;RS...}(x^{m},y)\rightarrow\Phi_{AB...}%
^{\;\;\;\;\;\;RS...}(x^{m},-y).
\end{equation}
Each $\left(  D-1\right)  $ index will induce an overall minus sign in the
fields $\Phi_{AB...}^{\;\;\;\;\;\;RS...}(x^{m},y)$ under this
transformation,\ thus making the corresponding $x$-dependent component to
become real when the number of indices with value $\left(  D-1\right)  $ is
even, and purely imaginary in the case it has an odd number of indices with
this value. Thus, for example, when the field has no indices with this value
we get
\begin{equation}
\Phi_{ab...}^{\;\;\;\;\;\;rs...}(x^{m},y)=\sqrt{\frac{\mu}{\pi}}\Phi
_{ab...}^{\;\;\;\;\;\;rs...}(x^{m})\cos\mu y.\label{EXPEVEN}%
\end{equation}
When the field has one index with this value, we denote such components by%
\begin{equation}
\Phi_{\left(  D-1\right)  b...}^{\;\;\;\;\;\;\ \ \ rs...}(x^{m})=-i\tilde
{\Phi}_{b...}^{\;\;\;\;\;\;rs...}(x^{m})
\end{equation}
where the tensor with a tilde has only $\left(  D-2\right)  $-dimensional
indices and it is real . In this way we write%
\begin{equation}
\Phi_{\left(  D-1\right)  b...}^{\;\;\;\;\;\;\ \ \ \ \ \ rs...}(x^{m}%
,y)=\sqrt{\frac{\mu}{\pi}}\tilde{\Phi}_{b...}^{\;\;\;\;rs...}(x^{m})\sin\mu
y.\label{EXPODD}%
\end{equation}
It is clear that the expressions (\ref{EXPEVEN})\ and ((\ref{EXPODD})) can be
generalized to any tensor having an even or odd number of subindexes $\left(
D-1\right)  $. In general we will use different names for these reduced
tensors, dropping the indices with value $\left(  D-1\right)  $.

After the $(a,D-1)$ separation has been made in the coordinates and fields,
the resulting four dimensional action is obtained by performing the
integration of $y$ over a circle. The only surviving contributions come from
\begin{equation}
\oint dy\;\cos^{2}\mu y=\oint dy\;\sin^{2}\mu y=\frac{\pi}{\mu}.
\end{equation}
In the sequel we denote any function $\Sigma(x^{m},y)$ by $\Sigma(x,y)$.

We will show now that this dimensional reduction applied to the $D$%
-dimensional massless spin two field, actually yields the massive
$(D-1)$-dimensional FP theory. We start from the action for the massless spin
two field in $D$ dimensions%
\begin{equation}
S_{D}=\frac{1}{2}\int d^{4}x\;dy\left(  -\partial_{A}e^{\left\{  MN\right\}
}\partial^{A}e_{\left\{  MN\right\}  }+2\partial_{M}e^{\left\{  MN\right\}
}\partial^{A}e_{\left\{  AN\right\}  }-2\partial_{M}e^{\left\{  MN\right\}
}\partial_{N}e+\partial_{A}e\partial^{A}e\right)  ,\label{FPL5M0}%
\end{equation}
with $e_{\left\{  MN\right\}  }=e_{\left\{  NM\right\}  }$, which is invariant
under the gauge transformations%
\begin{equation}
\delta e_{\left\{  MN\right\}  }=\partial_{M}\xi_{N}+\partial_{N}\xi
_{M}.\label{GT}%
\end{equation}
The dimensional reduction is implemented in term of the fields $e_{mn}(x)$,
$a_{m}(x)$, $\varphi(x)$ , defined by%
\begin{align}
e_{\left\{  mn\right\}  }(x,y)  &  =\sqrt{\frac{\mu}{\pi}}e_{\left\{
mn\right\}  }(x)\cos\mu y,\\
e_{\left\{  \left(  D-1\right)  n\right\}  }(x,y)  &  =\sqrt{\frac{\mu}{\pi}%
}a_{m}(x)\sin\mu y,\\
e_{\left\{  \left(  D-1\right)  \left(  D-1\right)  \right\}  }(x,y)  &
=\sqrt{\frac{\mu}{\pi}}\varphi(x)\cos\mu y,\label{FPF}%
\end{align}
while the gauge transformations (\ref{GT}) are translated into%
\begin{equation}
\delta e_{\left\{  mn\right\}  }=\partial_{m}\xi_{n}+\partial_{n}\xi
_{m},\;\;\delta a_{m}=\partial_{m}\xi-\mu\xi_{m},\;\;\delta\varphi=2\mu
\xi(x).\label{INDGT}%
\end{equation}
with
\begin{equation}
\xi_{m}(x,y)=\sqrt{\frac{\mu}{\pi}}\xi_{m}(x)\cos\mu y,\;\;\;\xi
_{(D-1)}(x,y)=\sqrt{\frac{\mu}{\pi}}\xi(x)\sin\mu y.
\end{equation}
Redefining
\begin{equation}
\bar{a}_{m}=a_{m}-\frac{1}{2\mu}\partial_{m}\varphi,\label{REDEFa}%
\end{equation}
the reduced action $\ $is
\begin{align}
S_{\left(  D-1\right)  }  &  =\frac{1}{2}\int d^{4}x\;\left\{  -\partial
_{a}e^{\left\{  mn\right\}  }\partial^{a}e_{\left\{  mn\right\}  }%
+2\partial_{m}e^{\left\{  mn\right\}  }\partial^{a}e_{\left\{  an\right\}
}-2\partial_{m}e^{\left\{  mn\right\}  }\partial_{n}e+\left(  \partial
_{a}e\right)  ^{2}\right. \nonumber\\
&  \left.  -\mu^{2}\left[  \left(  e_{\left\{  mn\right\}  }+\frac{1}{\mu
}\left(  \partial_{m}\bar{a}_{n}+\partial_{n}\bar{a}_{m}\right)  \right)
^{2}-\left(  e+\frac{2}{\mu}\partial^{n}\bar{a}_{n}\right)  ^{2}\right]
\right\}  ,\label{FPA4M}%
\end{align}
where $e=e_{m}^{\;\;\;m}$. The action (\ref{FPA4M})\ remains invariant under
the induced gauge transformations%
\begin{equation}
\delta e_{\left\{  mn\right\}  }=\partial_{m}\xi_{n}+\partial_{n}\xi
_{m},\;\;\delta\bar{a}_{n}=-\mu\xi_{n}.\label{FINGT}%
\end{equation}

To get the FP action we can now fix the gauge, choosing $\xi_{n}(x)$ such that
$\bar{a}_{n}=0$.\ This leaves us with $e_{\left\{  mn\right\}  }$ as the
remaining degrees of freedom, with the standard $(D-1)$-dimensional action%
\begin{equation}
S_{\left(  D-1\right)  }=\frac{1}{2}\int d^{4}x\ \left[  -\partial
_{a}e^{\left\{  mn\right\}  }\partial^{a}e_{\left\{  mn\right\}  }%
+2\partial_{m}e^{\left\{  mn\right\}  }\partial^{a}e_{\left\{  an\right\}
}-2\partial_{m}e^{\left\{  mn\right\}  }\partial_{n}e+\left(  \partial
_{a}e\right)  ^{2}-\mu^{2}\left(  e_{\left\{  mn\right\}  }e^{\left\{
mn\right\}  }-e^{2}\right)  \right]  .\label{FPA4MFIN}%
\end{equation}

\section{Dimensional reduction of the $D$ dimensional massless $s=2$ parent
action.}

It is well known that the first order parent action
\begin{equation}
S=-\frac{1}{2}\int d^{\left(  D-1\right)  }y\left[  Y^{C[AB]}\left(
\partial_{A}e_{BC}-\partial_{B}e_{AC}\right)  -Y_{C[AB]}Y^{B[AC]}+\frac
{1}{D-2}Y_{A}Y^{A}\right]  ,\;\;Y_{\;\;[AB]}^{B\;\;\;\ \ \ }=Y_{A}%
,\label{ACC5}%
\end{equation}
with $Y^{C[AB]}=-Y^{C[BA]}$, generates massless dual\ theories for the spin
two field in $D$ dimensions \cite{Boulanger3}. The field $Y^{C[AB]}$ has
$D^{2}(D-1)/2$ independent components while $e_{BC}\neq e_{CB}$ accounts for
$D^{2}$, which give a total of $D^{2}\left(  D+1\right)  /2$ independent
components. The above action is invariant under the gauge transformations
(local Lorentz transformations)%
\begin{align}
\delta Y^{C[AB]} &  =-\left[  \partial^{C}\omega^{\lbrack AB]}+\partial
_{D}\omega^{\lbrack BD]}\eta^{AC}+\partial_{D}\omega^{\lbrack DA]}\eta
^{BC}\right]  ,\\
\delta Y_{A} &  =-\left(  D-2\right)  \partial^{D}\omega_{\lbrack DA]},\qquad
Y_{A}=Y_{C[AB]}\eta^{BC},\\
\delta e_{BC} &  =\omega_{\lbrack BC]},
\end{align}
together with (local diffeomorphisms)%
\begin{align}
\delta Y^{D[AB]} &  =\partial^{D}\left(  \partial^{A}\xi^{B}-\partial^{B}%
\xi^{A}\right)  +\left(  \eta^{AD}\partial^{B}-\eta^{BD}\partial^{A}\right)
\partial_{C}\xi^{C}+\eta^{BD}\partial^{2}\xi^{A}-\eta^{AD}\partial^{2}\xi
^{B},\\
\delta Y^{A} &  =(D-2)\left(  \partial^{2}\xi^{A}-\partial^{A}\partial_{C}%
\xi^{C}\right)  ,\\
\delta e_{AB} &  =\partial_{A}\xi_{B}+\partial_{B}\xi_{A}.
\end{align}
According to Ref. \cite{Boulanger3}, these gauge symmetries are independent of
the number of dimensions.

The dimensional reduction is performed via the following redefinitions for the
fields%
\begin{align}
Y^{c[ab]}(x,y) &  =\sqrt{\frac{\mu}{\pi}}Y^{c[ab]}(x)\cos\mu y,\;\;Y^{\left(
D-1\right)  [b\left(  D-1\right)  ]}(x,y)=\sqrt{\frac{\mu}{\pi}}Z^{b}\cos\mu
y,\label{CAMP1}\\
Y^{\left(  D-1\right)  [ab]}(x,y) &  =\sqrt{\frac{\mu}{\pi}}V^{[ab]}\sin\mu
y,\;\;\ \text{\ }\ \;\;\;Y^{c[b\left(  D-1\right)  ]}(x,y)=\sqrt{\frac{\mu
}{\pi}}W^{bc}\sin\mu y,
\end{align}%
\begin{align}
e_{ab}(x,y) &  =\sqrt{\frac{\mu}{\pi}}e_{ab}(x)\cos\mu y,\;\;e_{\left(
D-1\right)  \left(  D-1\right)  }(x,y)=\sqrt{\frac{\mu}{\pi}}S(x)\cos\mu y,\\
e_{a\left(  D-1\right)  }(x,y) &  =\sqrt{\frac{\mu}{\pi}}B_{a}(x)\sin\mu
y,\;\;\;\;e_{\left(  D-1\right)  a}(x,y)=\sqrt{\frac{\mu}{\pi}}A_{a}(x)\sin\mu
y,\label{CAMP2}%
\end{align}
which reshuffles the original independent components in the following way%
\begin{equation}
Y^{C[AB]}\rightarrow D^{2}(D-1)/2\ \ \ \left\{
\begin{array}
[c]{c}%
Y^{c[ab]}\rightarrow\;\left[  (D-1)^{2}(D-2)/2\right]  ,\\
V^{[ab]}\rightarrow\;\left[  (D-1)(D-2)/2\right]  ,\\
W^{bc}\rightarrow\;\left[  (D-1)^{2}\right]  ,\\
Z^{b}\rightarrow\;\left[  D-1\right]
\end{array}
\right.  ,
\end{equation}%
\begin{equation}
e_{BC}\rightarrow D^{2}\ \ \ \left\{
\begin{array}
[c]{c}%
e_{ab}\rightarrow\;(D-1)^{2},\\
A_{a\;}\rightarrow\;(D-1),\\
B_{a}\rightarrow\;(D-1),\\
S\rightarrow\;1
\end{array}
\right.  .
\end{equation}

Also the $D(D+1)/2$ gauge parameters are reorganized according to%
\begin{align}
\omega_{\lbrack ab]}(x,y)  &  =\sqrt{\frac{\mu}{\pi}}\omega_{\lbrack
ab]}(x)\cos\mu y,\;\;\;\;\;\omega_{\lbrack\left(  D-1\right)  a]}%
(x,y)=\sqrt{\frac{\mu}{\pi}}\omega_{a}(x)\sin\mu y,\\
\xi_{a}(x,y)  &  =\sqrt{\frac{\mu}{\pi}}\xi_{a}(x)\cos\mu
y,\;\ \ \ \ \ \ \ \ \ \ \ \ \ \ \ \xi_{\left(  D-1\right)  }(x,y)=\sqrt
{\frac{\mu}{\pi}}\xi(x)\sin\mu y.
\end{align}

The corresponding gauge transformations in the $(D-1)$ dimensional fields
associated to the $\left(  D-1\right)  (D-2)/2$ parameters $\omega^{\lbrack
ab]}$ and the $\left(  D-1\right)  $ parameters $\omega_{a}$\ can be rewritten
as:%
\begin{align}
\delta e_{\{ab\}}(x) &  =0,\;\ \;\;\;\;\ \ \ \ \delta e_{[ab]}(x)=\omega
_{\lbrack ab]}(x),\label{TNW1}\\
\delta B_{a}(x) &  =-\omega_{a}(x),\;\;\delta A_{a}(x)=\omega_{a}%
(x),\;\;\delta S=0,\;\label{TNW11}%
\end{align}%
\begin{align}
\delta Y^{c[ab]}(x) &  =-\left(  \partial^{c}\omega^{\lbrack ab]}+\partial
_{m}\omega^{\lbrack bm]}\eta^{ac}+\partial_{m}\omega^{\lbrack ma]}\eta
^{bc}\right)  -\mu\left(  \omega^{a}\eta^{bc}-\omega^{b}\eta^{ac}\right)  ,\\
\delta Y_{\;\;\;\;\;\;}^{a}(x) &  =-\left(  \partial_{b}\omega^{\lbrack
ab]}+3\partial_{m}\omega^{\lbrack ma]}\right)  -3\mu\omega^{a},\label{TNW2}%
\end{align}%
\begin{align}
\delta V^{[ab]} &  =\mu\omega^{\lbrack ab]},\;\;\;\delta Z^{a}=-\partial
_{m}\omega^{\lbrack ma]},\\
\delta W^{ac} &  =\partial^{c}\omega^{a}-\partial_{m}\omega^{m}\eta
^{ac}.\label{TNW3}%
\end{align}
Let us notice that $\partial_{a}\delta W^{ac}=\partial^{c}\partial_{a}%
\omega^{a}-\partial^{c}\partial_{m}\omega^{m}=0$. The remaining gauge
transformations, given by the $(D-1)$ parameters\ $\xi^{a}$\ and the parameter
$\xi$ are:
\begin{align}
\delta e_{\{ab\}}(x) &  =\partial_{a}\xi_{b}+\partial_{b}\xi_{a}%
,\;\;\;\;\;\;\delta e_{[ab]}(x)=0,\\
\delta B_{a}(x) &  =\delta A_{a}(x)=\partial_{a}\xi-\mu\xi_{a},\;\;\delta
S=2\mu\xi,\label{TNXI1}%
\end{align}%
\begin{align}
\delta Y^{c[ab]}(x) &  =\partial^{c}\left(  \partial^{a}\xi^{b}-\partial
^{b}\xi^{a}\right)  +\left(  \eta^{ac}\partial^{b}-\eta^{bc}\partial
^{a}\right)  \partial_{m}\xi^{m}+\partial^{2}\left(  \eta^{bc}\xi^{a}%
-\eta^{ac}\xi^{b}\right) \nonumber\\
&  +\mu\left(  \eta^{ac}\partial^{b}-\eta^{bc}\partial^{a}\right)  \xi-\mu
^{2}\left(  \eta^{bc}\xi^{a}-\eta^{ac}\xi^{b}\right)  ,\label{TNXI2}\\
\delta Y^{a}(x) &  =2\partial^{2}\xi^{a}-2\partial^{a}\partial_{m}\xi^{m}%
-3\mu\partial^{a}\xi-3\mu^{2}\xi^{a},
\end{align}%
\begin{align}
\delta V^{[ab]} &  =-\mu\left(  \partial^{a}\xi^{b}-\partial^{b}\xi
^{a}\right)  ,\;\;\delta Z^{a}=\partial^{2}\xi^{a}-\partial^{a}\partial_{m}%
\xi^{m},\\
\delta W^{ad} &  =\partial^{d}\partial^{a}\xi-\eta^{ad}\bar{\partial}^{2}%
\xi+\mu\left(  \partial^{d}\xi^{a}-\eta^{ad}\partial_{m}\xi^{m}\right)
.\label{TNXI3}%
\end{align}
Again we have here $\partial_{a}\delta W^{ac}=0$. In the above $\partial
^{2}=-\partial_{0}^{2}+\mathbf{\nabla}^{2}$ denotes the $(D-1)$-dimensional D'Alambertian.

After substituting the fields (\ref{CAMP1}-\ref{CAMP2}) in the $D$ dimensional
action (\ref{ACC5}) and performing the integration with respect to the fifth
coordinate $y$ we obtain the following dimensionally reduced parent action in
$(D-1)$ dimensions%

\begin{align}
S &  =-\frac{1}{2}\int d^{4}x\left\{  Y^{c[ab]}\left(  \partial_{a}%
e_{bc}-\partial_{b}e_{ac}\right)  -Y_{c[ab]}Y^{b[ac]}+\frac{1}{D-2}Y_{a}%
Y^{a}\right.  \nonumber\\
&  +V^{[ab]}\left(  \partial_{a}B_{b}-\partial_{b}B_{a}\right)  +2W^{ac}%
\partial_{a}A_{c}+2\mu W^{ac}e_{ac}-2V_{[ab]}W^{ab}-W_{bc}W^{cb}\nonumber\\
&  \left.  +2Z^{a}\partial_{a}S-2\mu Z^{a}B_{a}+\frac{2}{D-2}Z_{a}Y^{a}%
-\frac{D-3}{D-2}Z_{a}Z^{a}+\frac{1}{D-2)}W^{2}\right\}  ,\label{DIMREDACT}%
\end{align}
where $W=W_{\ \ a}^{a}$.\ This parent action contains $D^{2}\left(
D+1\right)  /2$\ fields and $D(D+1)/2$\ arbitrary functions to be gauge fixed.
After the gauge fixing $D\left(  D^{2}-1\right)  /2$\ variables remain. Going
from the $D\left(  D^{2}-1\right)  /2$\ remaining variables to the final
$D(D-3)/2$\ degrees of freedom requires the elimination via equations of
motion of some of the remaining variables, which act as auxiliary fields. The
gauge fixed Lagrangians are equivalent in the sense that all of them are in a
gauge orbit, but the subsequent elimination of auxiliary fields depends on the
gauge fixing and breaks this equivalence.

According to the gauge transformations (\ref{TNW11}) and (\ref{TNXI1}), the
fields $A_{a}$, $B_{a}$ and $S$ are pure gauge fields and can be completely
fixed by an adequate choice of $\omega_{a}$, $\xi_{a}$ and $\xi$. With this
partial gauge fixing in the action (\ref{DIMREDACT}), $W_{bc}$, $V_{a}$ and
$Z_{a}$ are purely algebraic fields, and thus all the dynamics is contained in
the fields $Y^{c[ab]}$ and $e_{ac}$. The remaining gauge symmetry, related to
$\omega_{\lbrack ab]}$, can be used either to set zero the antisymmetric part
of $e_{ac}$, in which case $V^{[ab]}$ becomes a Lagrange multiplier for
$W_{bc}$, or to fix $V^{[ab]}$, in which case $e_{ac} $ has no definite
symmetry. These two possibilities are considered in the following section.

\section{GAUGE FIXING AND\ AUXILIARY VARIABLE\ ELIMINATION\ IN THE PARENT
ACTION}

Gauge invariance is preserved by the above dimensional reduction. In fact, we
have explicitly verified that the action (\ref{DIMREDACT}) is invariant under
the full set of gauge transformations (\ref{TNW1})-(\ref{TNXI3}). In this
sense, the action (\ref{DIMREDACT})\ is of the Stueckelberg type, being of
similar character than those obtained in Refs. \cite{ZINOVIEV1,ZINOVIEV2}. In
the following we explore the two gauge fixings mentioned at the end of the
preceding section, followed by the subsequent elimination of auxiliary variables.

\subsection{GAUGE \ FIXING LEADING TO A PARENT ACTION WITH SYMMETRICAL
$e_{\left\{  bc\right\}  }$}

In this case we fix the gauges by choosing the infinitesimal parameters
$\omega^{\lbrack ab]}$, $\omega_{a}$,\ $\xi_{a}$,\ $\xi$\ as follows. We have
the transformations \
\begin{align}
\bar{e}^{[ab]} &  =e^{[ab]}+\omega^{\lbrack ab]},\\
\bar{S} &  =S+2\mu\xi,\\
\bar{A}_{a} &  =A_{a}+\omega_{a}+\partial_{a}\xi-\mu\xi_{a},\\
\bar{B}_{a} &  =B_{a}-\omega_{a}+\partial_{a}\xi-\mu\xi_{a},
\end{align}
where we are temporarily denoting the gauge transformed fields by a bar. We
take $\omega^{\lbrack ab]}$ such that $\bar{e}^{[ab]}=0$, i.e. only the
symmetric part $\bar{e}^{\{ab\}}$ survives. Besides, we choose the remaining
parameters in such a way that
\begin{equation}
\bar{S}=\bar{A}_{a}=\bar{B}_{a}=0.
\end{equation}
This can be done by taking
\begin{equation}
\xi=-\frac{1}{2\mu}S,\;\omega_{a}=\frac{1}{2}\left(  B_{a}-A_{a}\right)
,\;\xi_{a}=\frac{1}{2\mu}\left(  A_{a}+B_{a}\right)  -\frac{1}{4\mu^{2}%
}\partial_{a}S.
\end{equation}
Thus the gauge fixed parent action becomes%
\begin{align}
S &  =-\frac{1}{2}\int d^{4}x\;\left\{  Y^{c[ab]}\left(  \partial
_{a}e_{\left\{  bc\right\}  }-\partial_{b}e_{\left\{  ac\right\}  }\right)
+2\mu W^{ac}e_{\left\{  ac\right\}  }\right. \nonumber\\
&  -Y_{c[ab]}Y^{b[ac]}+\frac{1}{D-2}Y_{a}Y^{a}-W_{bc}W^{cb}+\frac{1}{D-2}%
W^{2}-\frac{D-3}{D-2}Z_{a}Z^{a}\nonumber\\
&  \left.  -2W^{ab}V_{[ab]}+\frac{2}{(D-2)}Z_{a}Y^{a}\right\}
.\label{DM1PGFA}%
\end{align}
where the bars of the gauge transformed fields have been dropped. Here
$V_{[ac]}$ acts as a Lagrange multiplier that produces the constraint
\begin{equation}
W^{[ac]}=0,
\end{equation}
which is immediately implemented by just leaving the symmetric part of
$W^{ab}$, $W^{\left\{  ab\right\}  }$, in the action. We still have some
auxiliary fields that can be eliminated from the action. They are $W^{\{ab\}}
$ itself and $Z_{a}$, which are algebraically determined by their equations of
motion%
\begin{align}
Z^{a} &  =\frac{1}{(D-3)}Y^{a},\\
W_{\{bc\}} &  =\mu\left(  e_{\{bc\}}-\eta_{bc}e\right)  ,\ \ \ W=-\left(
D-2\right)  \mu e.
\end{align}

Substituting in (\ref{DM1PGFA}), our final expression for the $(D-1)$%
-dimensionally reduced massive parent action is%
\begin{equation}
S=-\frac{1}{2}\int d^{4}x\left[  Y^{c[ab]}\left(  \partial_{a}e_{\left\{
bc\right\}  }-\partial_{b}e_{\left\{  ac\right\}  }\right)  -Y_{[cab]}%
Y^{b[ac]}+\frac{1}{D-3}Y_{a}Y^{a}+\mu^{2}\left(  e_{\left\{  bc\right\}
}e^{\left\{  bc\right\}  }-e^{2}\right)  \right]  ,\label{ACCM}%
\end{equation}
with $e=e_{\{ab\}}\eta^{ab}$. By eliminating $Y^{[ab]c}$ we recover the FP
action, and the elimination of $e_{\left\{  bc\right\}  }$ leads to a dual
action of the form discussed in Ref. \cite{CMU1} for $D=4$ dimensions.

\subsection{GAUGE FIXING LEADING TO A PARENT ACTION WITH A NON-SYMMETRICAL
$e_{bc}$}

Next we apply a gauge fixing partially similar to the one of the preceding
section. We still fix the gauge in such a way that%
\begin{equation}
\bar{S}=0=\bar{A}_{a}=0=\bar{B}_{a},
\end{equation}
but the gauge freedom in$\;\omega^{\left[  ab\right]  }$ is used to set
\[
V^{[ab]}=0,
\]
instead of $e_{\left[  bc\right]  }=0$, and thus $e_{bc}$ have no definite
symmetry. The parent action results in%
\begin{align}
S &  =-\frac{1}{2}\int d^{4}x\left\{  Y^{c[ab]}\left(  \partial_{a}%
e_{bc}-\partial_{b}e_{ac}\right)  -Y_{c[ab]}Y^{b[ac]}+\frac{1}{D-2}Y_{a}%
Y^{a}\right.  \nonumber\\
&  +2\mu W^{ac}e_{ac}-W_{bc}W^{cb}+\frac{1}{D-2}W^{2}\nonumber\\
&  \left.  +\frac{2}{D-2}Z_{a}Y^{a}-\frac{D-3}{D-2}Z_{a}Z^{a}\right\}
.\label{DM2PGFA}%
\end{align}
Note that $W^{bc}$ is not constrained to have a definite symmetry. As before,
we eliminate $Z_{a}$ and $W^{bc}$ using the corresponding equations of motion.
The case of $Z_{a}$ is the same as in the previous section so that we obtain%
\begin{align}
S &  =-\frac{1}{2}\int d^{4}x\left\{  Y^{c[ab]}\left(  \partial_{a}%
e_{bc}-\partial_{b}e_{ac}\right)  -Y_{c[ab]}Y^{b[ac]}+\frac{1}{D-3}Y_{a}%
Y^{a}\right.  \nonumber\\
&  \left.  +2\mu W^{ac}e_{ac}-W_{bc}W^{cb}+\frac{1}{D-2}W^{2}\right\}  .
\end{align}
Next, the elimination of $W^{bc}$ produces%
\[
W_{cb}=\mu\left(  e_{bc}-\eta_{bc}e\right)  ,\;\;\;\;W=-(D-2)\mu e.
\]
Finally we get
\begin{equation}
S=-\frac{1}{2}\int d^{4}x\left[  Y^{c[ab]}\left(  \partial_{a}e_{bc}%
-\partial_{b}e_{ac}\right)  -Y_{c[ab]}Y^{b[ac]}+\frac{1}{D-3}Y_{a}Y^{a}%
+\mu^{2}\left(  e^{bc}e_{cb}-e^{2}\right)  \right]  .
\end{equation}
as the final parent action in this sequence of gauge fixings and field
eliminations, which is analogous to the one obtained in the preceding section,
but with $e_{bc}$ without a definite symmetry. This is precisely the action
obtained in Ref. \cite{West}.

The gauge fixed actions (\ref{DM1PGFA}) and (\ref{DM2PGFA}) are equivalent in
the usual sense of gauge theories, but in each case the additional elimination
of auxiliary variables follows a different pattern. For this reason, although
both parent actions lead to the same action for $e_{\left\{  bc\right\}  }$
after eliminating $Y^{c[ab]}$, they yield different dual theories after
eliminating either $e_{\left\{  bc\right\}  }$ or $e_{bc}$. The case discussed
in this subsection reproduces the Curtright-Freund \cite{CURT} dual theory
when restricted to $D=4$.

\section{DUAL\ THEORIES}

In this section we show that the two sequences of gauge fixings
and field eliminations proposed above lead to the standard FP theory
on one hand, but to completely different dual actions on the other. Once we
have obtained the massive parent action from dimensional reduction we set
$\left(  D-1\right)  $ to $D$ and relabel the tensor indices with capital letters.

\subsection{THE CASE OF A SYMMETRICAL $e_{\left\{  BC\right\}  }$}

In a flat $D$-dimensional space-time we take
\begin{equation}
S=-\frac{1}{2}\int d^{D}x\left[  Y^{C[AB]}\left(  \partial_{A}e_{\left\{
BC\right\}  }-\partial_{B}e_{\left\{  AC\right\}  }\right)  +Y_{C[AB]}%
Y^{B[AC]}-\frac{1}{(D-2)}Y_{A}Y^{A}-\mu^{2}\left(  e_{\left\{  AB\right\}
}e^{\left\{  AB\right\}  }-e^{2}\right)  \right]  ,\label{PACTDDIM}%
\end{equation}
as our parent action. Here the fields are $e_{\left\{  BC\right\}  }=$
$e_{\left\{  CB\right\}  }$ and $Y^{C[AB]}=-Y^{C[BA]}$, which have $D(D+1)/2$
and $D^{2}(D-1)/2$ components respectively.

Eliminating $Y^{C[AB]}$ using its Euler-Lagrange equations%
\begin{align}
Y_{C[AB]} &  =-\left(  \partial_{A}e_{BC}-\partial_{B}e_{AC}\right)  +\left(
\partial_{A}e-\partial^{M}e_{AM}\right)  \eta_{BC}-\left(  \partial
_{B}e-\partial^{M}e_{BM}\right)  \eta_{AC},\\
Y_{A} &  =\left(  D-2\right)  \left(  \partial_{A}e-\partial^{B}e_{AB}\right)
,\label{YH}%
\end{align}
yields finally to
\begin{equation}
S=\frac{1}{2}\int d^{D}x\left[  -\partial_{C}e_{BA}\partial^{C}e^{BA}%
+\partial_{C}e\partial^{C}e-2\partial^{M}e\partial^{N}e_{NM}+2\partial
^{M}e_{MA}\partial_{N}e^{NA}-\mu^{2}\left(  e_{AB}e^{AB}-e^{2}\right)
\right]  ,\label{FPDIMD}%
\end{equation}
which is precisely the FP action in $D$ dimensions.

To obtain the dual action we eliminate $e^{\left\{  BA\right\}  }$ from its
equations of motion obtained from\ (\ref{PACTDDIM}). It is convenient to
introduce the decomposition
\begin{equation}
Y_{R\left[  PQ\right]  }=C_{R\left[  PQ\right]  }+A_{\left[  PQR\right]
},\label{DECOMP}%
\end{equation}
where the field $A_{\left[  PQR\right]  }$, which has $D(D-1)(D-2)/6$
independent components, is completely antisymmetric in all indices, while
$C_{R\left[  PQ\right]  }$ satisfies the cyclic identity%
\begin{equation}
C_{R\left[  PQ\right]  }+C_{P\left[  QR\right]  }+C_{Q\left[  RP\right]
}=0,\;\;\;\longleftrightarrow\;\;C_{C\left[  AB\right]  }\epsilon
^{ABCN_{4}...N_{D}}=0.\label{CYCLID}%
\end{equation}
This splitting works because the number of constraints arising from the cyclic
identity is precisely $D(D-1)(D-2)/6$.

In terms of this new field, the action (\ref{PACTDDIM})\ results
\begin{equation}
S=\frac{1}{2}\int d^{D}x\left[  \left(  \partial^{C}e^{BA}-\partial^{B}%
e^{CA}\right)  C_{A\left[  CB\right]  }+\frac{1}{2}C_{C\left[  AB\right]
}C^{C\left[  AB\right]  }-A_{\left[  ABC\right]  }A^{\left[  ABC\right]
}-\frac{1}{(D-2)}C_{A}C^{A}-\mu^{2}\left(  e_{AB}e^{AB}-e^{2}\right)  \right]
,\label{MODPACT1}%
\end{equation}
where $C^{A}=C_{B\;\;\;\;\;\;}^{\;\;\left[  AB\right]  }.\;$In the above we
have used the cyclic identity to rewrite the quadratic terms in $C_{C\left[
AB\right]  }$ in the form $C_{C\left[  AB\right]  }C^{C\left[  AB\right]  }$.
The field $A_{\left[  ABC\right]  }$ decouples, leading to $A_{\left[
ABC\right]  }=0$ in virtue of its equations of motion. In order to make future
contact with Refs. \cite{CMU1, CMU2} we introduce the Hodge-dual of
$C^{C\left[  AB\right]  }$%
\begin{equation}
T_{P\left[  Q_{1}Q_{2}...Q_{D-2}\right]  }=\frac{1}{2}C_{P}^{\;\;\left[
AB\right]  }{}\epsilon_{ABQ_{1}Q_{2}...Q_{D-2}},
\end{equation}
which is a tensor of rank $\left(  D-1\right)  $ completely antisymmetric in
its last $\left(  D-2\right)  $ indices. The resulting action corresponding to
the field $T_{P\left[  Q_{1}Q_{2}...Q_{D-2}\right]  }$ will be taken as the
dual version of the original FP formulation. We can invert (\ref{DUALF})
obtaining%
\begin{align}
C_{P}^{\;\;\left[  AB\right]  } &  =-\frac{1}{(D-2)!}T_{P\left[  Q_{1}%
Q_{2}...Q_{D-2}\right]  }\epsilon^{Q_{1}Q_{2}...Q_{D-2}AB},\\
C^{A} &  =-\frac{1}{(D-2)!}\epsilon^{Q_{1}Q_{2}...Q_{D-2}AS}T_{S\left[
Q_{1}Q_{2}...Q_{D-2}\right]  }.
\end{align}
Notice that the cyclic identity of $C^{P\left[  AB\right]  }$ leads to the
traceless condition%
\begin{equation}
T_{\;\;\left[  PQ_{1}...Q_{D-3}\right]  }^{P}=0.\label{TRACELESS}%
\end{equation}
Let us remark that the kinetic part of the action for the field $T_{P\left[
Q_{1}Q_{2}...Q_{D-2}\right]  }$ will arise from the terms containing
$e_{\left\{  AB\right\}  }$ in (\ref{MODPACT1}), while the corresponding mass
terms are contained in the remaining pieces with the field $C_{A\left[
CB\right]  }$. In other words $S=S_{KIN}+S_{MASS}$, with%
\begin{equation}
S_{KIN}(e,T)=\frac{1}{2}\int d^{D}x\left[  2\partial^{C}e^{\left\{
BA\right\}  }C_{\left[  CB\right]  A}-\mu^{2}\left[  e_{\left\{  AB\right\}
}e^{\left\{  AB\right\}  }-e^{2}\right]  \right]  ,\label{SKIN}%
\end{equation}%
\begin{equation}
S_{MASS}(T)=\frac{1}{2}\int d^{D}x\left[  \frac{1}{2}C_{C\left[  AB\right]
}C^{C\left[  AB\right]  }-\frac{1}{D-2}C_{A}C^{A}\right]  .\label{SMASS}%
\end{equation}
The mass contribution produces
\begin{equation}
S_{MASS}(T)=-\frac{1}{2(D-2)!}\int d^{D}x\left[  \frac{D-3}{D-2}T_{A\left[
Q_{1}Q_{2}...Q_{D-2}\right]  }T^{A\left[  Q_{1}Q_{2}...Q_{D-2}\right]
}+T^{A\left[  BM_{1}M_{2}...M_{D-2}\right]  }T_{B\left[  AM_{1}M_{2}%
...M_{D-2}\right]  }\right]  .\label{SMASST}%
\end{equation}
The calculation of the kinetic contribution requires the equations of motion
for $e_{\left\{  AB\right\}  }$. Here it is convenient to introduce the field
strength $F^{B\left[  Q_{1}Q_{2}...Q_{D-2}Q_{D-1}\right]  }$, which is a
tensor of rank $D$, associated with the potential $T^{A\left[  Q_{1}%
Q_{2}...Q_{D-2}\right]  }$, given by%
\begin{equation}
F^{A\left[  Q_{1}Q_{2}...Q_{D-2}Q_{D-1}\right]  }=\frac{1}{\left(  D-2\right)
!}\delta_{\left[  A_{1}A_{2}...A_{D-2}A_{D-1}\right]  }^{\left[  Q_{1}%
Q_{2}...Q_{D-2}Q_{D-1}\right]  }\partial^{A_{1}}T^{A\left[  A_{2}%
...A_{D-2}A_{D-1}\right]  },\label{FIELDS}%
\end{equation}
which is completely antisymmetric with respect to the $(D-1)$ indices inside
the square brackets. Here $\delta_{\left[  A_{1}A_{2}...A_{D-2}A_{D-1}\right]
}^{\left[  Q_{1}Q_{2}...Q_{D-2}Q_{D-1}\right]  }$ denotes the completely
antisymmetric delta symbol. In this way $F^{A\left[  Q_{1}Q_{2}...Q_{D-2}%
Q_{D-1}\right]  }$ satisfies
\begin{equation}
\epsilon_{Q_{1}Q_{2}...Q_{D-2}Q_{D-1}B}\;F^{A\left[  Q_{D-1}Q_{1}%
Q_{2}...Q_{D-2}\right]  }=\left(  D-1\right)  !\;\epsilon_{Q_{1}%
Q_{2}...Q_{D-2}Q_{D-1}B}\partial^{Q_{D-1}}T^{\;A\left[  Q_{1}Q_{2}%
...Q_{D-2}\right]  }.
\end{equation}
In terms of the field strength the equations of motion for $e_{AB}$ lead to%
\begin{align}
e_{\left\{  AB\right\}  } &  =\frac{1}{2\mu^{2}\left(  D-1\right)  !}\;\left[
\epsilon_{Q_{1}Q_{2}...Q_{D-2}Q_{D-1}B}F_{A}^{\;\;\left[  Q_{D-1}Q_{1}%
Q_{2}...Q_{D-2}\right]  }+\epsilon_{Q_{1}Q_{2}...Q_{D-2}Q_{D-1}A}%
F_{B}^{\;\;\left[  Q_{D-1}Q_{1}Q_{2}...Q_{D-2}\right]  }\right.  \nonumber\\
&  \ \ \ \ \ \ \ \ \ \ \ \ \ \ \ \ \ \ \ \ \ \ \ \ \ \ \ \ \left.  -\frac
{2}{(D-1)}\epsilon_{Q_{1}Q_{2}...Q_{D-2}Q_{D-1}E}F^{E\left[  Q_{D-1}Q_{1}%
Q_{2}...Q_{D-2}\right]  }\;\eta_{AB}\right]  ,\label{HAB}%
\end{align}%
\begin{equation}
e=-\frac{1}{\mu^{2}(D-1)\left(  D-1\right)  !}\;\epsilon_{Q_{1}Q_{2}%
...Q_{D-2}Q_{D-1}B}F^{B\left[  Q_{D-1}Q_{1}Q_{2}...Q_{D-2}\right]  }.\label{H}%
\end{equation}
Using the field strength we can rewrite the coupling term in (\ref{SKIN}) as
\begin{equation}
\int d^{D}x\left[  \partial^{C}e^{BA}C_{A\left[  CB\right]  }\right]
=\frac{1}{(D-1)!}\int d^{D}x\;e_{\;\;A}^{B}\epsilon_{Q_{1}Q_{2}...Q_{D-2}%
Q_{D-1}B}\;F^{A\left[  Q_{1}Q_{2}...Q_{D-2}Q_{D-1}\right]  }.
\end{equation}

The expressions (\ref{HAB}) and (\ref{H}) \ imply that
\begin{equation}
\mu^{2}(e_{\left\{  AB\right\}  }-\eta_{AB}e)=\frac{1}{2(D-1)!}\epsilon
_{Q_{1}Q_{2}...Q_{D-2}Q_{D-1}B}\;F_{A}^{\;\;\left[  Q_{1}Q_{2}...Q_{D-2}%
Q_{D-1}\right]  }+(A\longleftrightarrow B),
\end{equation}
which allows us to rewrite the kinetic piece of the action (\ref{SKIN}) in the
convenient form%
\begin{equation}
S_{KIN}=\frac{\mu^{2}}{2}\int d^{D}x\;\;\left(  e_{\left\{  AB\right\}
}e^{\left\{  AB\right\}  }-e^{2}\right)  ,
\end{equation}
where we finally substitute the expressions of $e_{\left\{  AB\right\}  }$ as
functions of $F_{B}^{\;\;\left[  Q_{1}Q_{2}...Q_{D-2}Q_{D-1}\right]  }$. The
result is%
\begin{align}
S_{KIN}  &  =-\frac{\left(  D-2\right)  }{2\mu^{2}\left(  D-1\right)
!(D-1)}\int d^{D}x\;\left\{  F^{A\left[  Q_{1}Q_{2}...Q_{D-2}Q_{D-1}\right]
}\;F_{A\left[  Q_{1}Q_{2}...Q_{D-2}Q_{D-1}\right]  \;}\right. \nonumber\\
&  \ \ \ \ \ \ \ \ \ \ \ \ \ \ \ \ \ \ \ \ \ \ \ \ \ \ \ \ \ \ \ \ -\frac
{1}{2}\frac{(D-1)^{2}}{\left(  D-2\;\right)  }\;F_{A}^{\;\;\left[
AQ_{1}...Q_{D-3}Q_{D-2}\right]  }\;F_{_{\;\;\;\;\left[  BQ_{1}...Q_{D-3}%
Q_{D-2}\right]  }}^{B}\nonumber\\
&  \ \ \ \ \ \ \ \ \ \ \ \ \ \ \ \ \ \ \ \ \ \ \ \ \ \ \ \ \ \ \ \left.
\ +\frac{\left(  D-1\right)  }{\left(  D-2\right)  }\left[  F^{M\left[
NM_{3}...M_{D}\right]  }F_{N\left[  MM_{3}...M_{D}\right]  }\right]  \right\}
.\label{SKINT}%
\end{align}
The final action, dual to FP in arbitrary dimensions, is then%
\begin{align}
S(T)  &  =-\int d^{D}x\;\frac{(D-1)^{2}}{\left(  D-2\right)  }\left\{
\frac{\left(  D-2\right)  }{(D-1)^{2}}F^{\left[  Q_{1}Q_{2}...Q_{D-2}%
Q_{D-1}\right]  A}\;F_{\left[  Q_{1}Q_{2}...Q_{D-2}Q_{D-1}\right]  A\;}\right.
\nonumber\\
&  \ \ \ \ \ \ \ \ \ \ \ \ \ \ +\frac{1}{\left(  D-1\right)  }F^{B\left[
AM_{3}...M_{D}\right]  }F_{A\left[  BM_{3}...M_{D}\right]  }-\frac{1}{2}%
F_{A}^{\;\;\left[  Q_{1}...Q_{D-3}Q_{D-2}A\right]  }\;F_{_{\;\;\left[
Q_{1}...Q_{D-3}Q_{D-2}B\right]  }}^{B}\nonumber\\
&  \ \ \ \ \ \ \ \ \ \ \ \ \ \ \left.  +\mu^{2}\left(  \frac{\left(
D-3\right)  }{\left(  D-2\right)  }T_{A\left[  Q_{1}Q_{2}...Q_{D-2}\right]
}^{\;\;\;\;\;\;\;\;\;\;\;\;\;\;\;\;\;}T^{A\left[  Q_{1}Q_{2}...Q_{D-2}\right]
}+T^{A\left[  BM_{1}M_{2}...M_{D-3}\right]  }T_{B\left[  AM_{1}M_{2}%
...M_{D-3}\right]  }\right)  \right\}  ,\label{DUALSYM}%
\end{align}
where the original action has been adequately rescaled. Setting $D=4$ in the
above action leads to the case $a=e^{2}$ of the general Lagrangian (61) in
Ref. \cite{CMU2}.

\subsection{THE CASE OF A NON-SYMMETRICAL $e_{AC}$}

This case is discussed in full detail in Ref. \cite{GKMU} so that we only
recall the results here. The starting point here is the parent action%
\begin{align}
S &  =\frac{1}{2}\int d^{D}x\left\{  Y^{C[AB]}\left(  \partial_{B}%
e_{AC}-\partial_{A}e_{BC}\right)  -Y_{C[AB]}Y^{B[AC]}+\frac{1}{(D-2)}%
Y_{A}Y^{A}\right. \nonumber\\
&  \left.  +\mu^{2}\left(  e_{AB}e^{BA}-e^{2}\right)  \right\}
.\label{PACTDDIM1}%
\end{align}
Here the basic fields are the non-symmetrical $e_{BC}$ together with
$Y^{C[AB]}=-Y^{C[BA]}$ , with $D^{2}$ and $\ D^{2}(D-1)/2$ independent
components respectively. As shown in reference \cite{Boulanger3}, the above
Lagrangian in the massless case leads to the FP action, in terms of
$e_{\left\{  BC\right\}  }$ only, after $Y^{B\left[  AC\right]  }$ is
eliminated via the equations of motion. The massive case is completely
analogous because the equations of motion for $Y_{C\left[  AB\right]  }$ do
not involve the mass term \cite{GKMU}. Thus, the kinetic energy piece of the
action in terms of $e_{AB}$ involves the antisymmetric part $e_{\left[
AB\right]  }$ only as a total derivative. The mass term contributes with a
term proportional to $e_{\left[  AB\right]  }e^{\left[  AB\right]  }$, which
leads to the equation of motion $e_{\left[  AB\right]  }=0$. It is rather
remarkable that the FP formulation is recovered in spite that $e_{AB}$\ is non-symmetrical.

To obtain the dual description we eliminate $e^{BA}$ using the equations of
motion obtained from the action (\ref{PACTDDIM1}), leading to the following
action for $Y_{[AB]C}$
\begin{equation}
\mu^{2}S=\int d^{D}x\left[  \partial_{A}Y^{C[AB]}\partial^{E}Y_{B[EC]}%
-\frac{1}{D-1}(\partial_{A}Y^{A})^{2}+\mu^{2}\left(  Y_{C[AB]}Y^{B[AC]}%
-\frac{1}{D-2}Y_{A}Y^{A}\right)  \right]  .\label{cm5}%
\end{equation}
Next we implement the change of variables
\begin{equation}
Y^{C[AB]}=\bar{w}^{C[AB]}+\frac{1}{(D-1)}(\eta^{CB}Y^{A}-\eta^{CA}%
Y^{B}),\label{25tilde}%
\end{equation}
where $\bar{w}^{C[AB]}$ has a null trace, $\bar{w}^{B}=\bar{w}_{A}%
^{\ \ [AB]}=0$, and obtain%
\begin{equation}
S=\frac{1}{2}\int d^{D}x\left[  \partial_{A}\bar{w}^{C[BA]}\partial^{E}\bar
{w}_{B[CE]}+\mu^{2}\left(  \bar{w}^{C[BA]}\bar{w}_{B[CA]}-\frac{1}%
{(D-1)\left(  D-2\right)  }Y^{A}Y_{A}\right)  \right]  ,
\end{equation}
which clearly shows that the trace of $Y^{C[BA]}$ is an irrelevant variable
that can be eliminated from the Lagrangian using its equation of motion. Thus
we finally get%
\begin{equation}
S=\frac{1}{2}\int d^{D}x\left(  \partial_{A}\bar{w}^{C[AB]}\partial^{E}\bar
{w}_{B[EC]}+\mu^{2}\bar{w}^{C[AB]}\bar{w}_{A[CB]}\right)  .\label{LAGOMEGA}%
\end{equation}
Now we introduce the Hodge-dual of $\bar{w}^{C\left[  AB\right]  }$%
\begin{equation}
T_{P\left[  Q_{1}Q_{2}...Q_{D-2}\right]  }=\frac{1}{2}\bar{w}_{P}{}^{\left[
AB\right]  }\,\epsilon_{ABQ_{1}Q_{2}...Q_{D-2}},\label{DUALF}%
\end{equation}
which is a dimension-dependent tensor of rank $\left(  D-1\right)  $
completely antisymmetric in its last $\left(  D-2\right)  $ indices. The
resulting action corresponding to the field $T_{P\left[  Q_{1}Q_{2}%
...Q_{D-2}\right]  }$ will be taken as the dual version of the original FP
formulation.\ Finally we obtain%

\begin{align}
S(T)=-\int d^{D}x &  \left\{  \left[  \frac{1}{\left(  D-1\right)  }%
F_{B}^{\ \ \left[  AQ_{1}..Q_{D-2}\right]  }\;F_{\ \ \left[  AQ_{1}%
..Q_{D-2}\right]  }^{B}-F_{A}^{\ \ \left[  AQ_{1}..Q_{D-2}\right]
}F_{\ \ \left[  BQ_{1}..Q_{D-2}\right]  }^{B}\right]  \right. \nonumber\\
&  \left.  +\mu^{2}\left[  T_{B\left[  Q_{1}..Q_{D-2}\right]  }T^{B\left[
Q_{1}..Q_{D-2}\right]  }-(D-2)T_{\ \ \;\left[  CQ_{2}...Q_{D-3}\right]  }%
^{C}T_{B}^{\ \ \;\left[  BQ_{2}...Q_{D-3}\right]  }\right]  \right\}
,\label{SFINAL}%
\end{align}
with an adequate rescaling of the original action. The field strength
$F_{\ \ \left[  AQ_{1}..Q_{D-2}\right]  }^{B}$ has been already introduced in
Eq. (\ref{FIELDS}). The field $T_{B\left[  Q_{1}Q_{2}...Q_{D-2}\right]  }$
satisfies the cyclic condition%
\begin{equation}
\epsilon^{ASQ_{1}Q_{2}...Q_{D-2}}T_{S\left[  Q_{1}Q_{2}...Q_{D-2}\right]
}=0.\label{c1}%
\end{equation}
The action (\ref{SFINAL}) reduces to the Curtright-Freund action in four
dimensions.

The equations of motion are%

\begin{align}
&  \left[  \left(  D-2\right)  !\delta_{\left[  A_{1}...A_{D-1}\right]
}^{\left[  M_{1}..M_{D-1}\right]  }\delta_{C}^{B}-\delta_{\left[  A_{1}%
A_{2}...A_{D-1}\right]  }^{\left[  BQ_{2}...Q_{D-1}\right]  }\delta_{\left[
CQ_{2}..\ Q_{D-1}\right]  }^{\left[  M_{1}...M_{D-1}\right]  }\right]
\partial^{A_{1}}\partial_{M_{1}}T_{\ \ \ \left[  M_{2}..M_{D-1}\right]  }%
^{C}\nonumber\\
&  -\mu^{2}\left[  \left(  D-2\right)  !\right]  ^{2}\left[  T_{\ \ \ \left[
A_{2}...A_{D-1}\right]  }^{B}-\frac{1}{(D-3)!}\delta_{\left[  A_{2}%
.......A_{D-1}\right]  }^{\left[  BM_{3}...M_{D-1}\right]  }T_{\ \ \left[
CM_{3}...M_{D-1}\right]  }^{C}\right]  =0\label{tmov}%
\end{align}
and they imply that the field $T_{\ \ \ \left[  A_{2}...A_{D-1}\right]  }^{B}$
satisfies the additional constraints \cite{GKMU}
\begin{align}
T_{\ \ \ \left[  BA_{3}...A_{D-1}\right]  }^{B} &  =0,\label{c2}\\
\partial^{D}T_{\ \ \ \left[  DA_{3}...A_{D-1}\right]  }^{B} &  =0,\label{c3}\\
\partial_{B}T_{\ \ \ \left[  A_{2}A_{3}...A_{D-1}\right]  }^{B} &
=0.\label{c4}%
\end{align}
After implementing these constraints the equation of motion reduces to its
simplest form%
\begin{equation}
\left(  \partial^{2}-\mu^{2}\right)  T_{\ \ \ \left[  A_{2}...A_{D-1}\right]
}^{B}=0.
\end{equation}

The field $T_{\ \ \ \left[  A_{2}...A_{D-1}\right]  }^{B}$ in $D$ dimensions
has $I=D^{2}(D-1)/2$ components, but the constraints (\ref{c1},\ref{c2}%
-\ref{c4}) manage to leave just $\frac{1}{2}D\left(  D-1\right)  -1$
independent degrees of freedom, which indeed is the same number obtained for
$e_{\left\{  AB\right\}  }$ in the FP formulation.

\section{FINAL COMMENTS}

In this paper we have explored a dimensional reduction from $D$ to $(D-1)$
dimensions in order to produce dual theories for massive spin two fields using
the parent action method. We started from the corresponding massless action in
the higher dimension and generated the mass parameter via dimensional
reduction, thus obtaining a lower dimension massive parent action. The massive
parent theory inherits all the gauge symmetries of the parent massless action,
so that it becomes a Stueckelberg-like action in dimension $\left(
D-1\right)  $. Although this parent action contains several fields, the
degrees of freedom are only contained in two of them, $Y^{C[AB]}$ and $e_{AB}%
$. Even so, the existence of alternative gauge choices together with
alternative auxiliary field eliminations via their equations of motion allowed
us to identify two kinds of $\left(  D-1\right)  $ dimensional massive parent
actions, corresponding either to a symmetric or a non-symmetric standard spin
two field $e_{AB}$. The true degrees of freedom for the resulting Fierz-Pauli
theory in terms of the field $e_{AB}$ are contained only in the symmetric
piece $e_{\left\{  AB\right\}  }$, in analogy to the massless case
\cite{Boulanger3,West}. Nevertheless, important differences arose in the
corresponding dual theories. In both cases we constructed the dual theory in
terms of a mixed symmetry field $T_{A\left[  B_{1}B_{2}...B_{D-2}\right]  }$
which final action is written without the use of auxiliary fields. The general
results are given in Eqs. (\ref{DUALSYM}) and (\ref{SFINAL}), respectively.
Let us emphasize that in both cases the dual theory to Fierz-Pauli is
constructed in terms of the $(D-1)$-rank tensor $T_{A\left[  B_{1}%
B_{2}...B_{D-2}\right]  }$, but subjected either to a traceless condition or
to a cyclic identity. Let us recall that in the massless case the dual to the
Fierz-Pauli field $e_{\left\{  AB\right\}  }$ is the $(D-2)$-rank tensor
$T_{A\left[  Q_{1}Q_{2}...Q_{D-3}\right]  }$ \cite{Boulanger3}. Notice that
this result is analogous to the well known one involving $p$-forms, where the
dual fields are a $(D-p-1)$-form for the massive case and a $(D-p-2)$-form for
the massless case. In the case of $D=4$ the symmetric case leads to a
particular family of dual actions previously found in Ref. \cite{CMU2}. The
non-symmetrical case reproduces the dual action proposed by Curtright and
Freund in Ref.\ \cite{CURT} . This constitutes the first proof that this
action is indeed dual to Fierz-Pauli. Finally, as a consistency check of our
procedure, we have considered in the Appendix the case $\mu=0$ in the $D=4$
parent action (\ref{DIMREDACT}). In this case, via adequate gauge fixings and
field eliminations, we recover the sum of the free spin two, one and zero
massless actions as expected, making up the original five degrees of freedom
we started with. One of the choices provides an unexpected Stueckelberg-like
formulation of the massless spin one field.

\appendix

\section{A FOUR DIMENSIONAL EXAMPLE}

Let us consider the massless ($\mu=0$)\ parent action (\ref{DIMREDACT}) in
four dimensions. Since we started from the massless spin two field in five
dimensions, which has five independent degrees of freedom we should be able to
select the corresponding gauge fixings and field eliminations in such a way to
recover the description of uncoupled massless fields of spins $2,1$ and
$0,$thus providing an alternative way of describing the five original degrees
of freedom. In the process we will be lead to a rather unexpected way of
presenting an action for the massless spin $1$ field, in terms of symmetric tensors.

We start from
\begin{align}
S &  =\frac{1}{2}\int d^{4}x\left\{  Y^{c[ab]}\left(  \partial_{a}%
e_{bc}-\partial_{b}e_{ac}\right)  -Y_{c[ab]}Y^{b[ac]}+\frac{1}{3}Y_{a}%
Y^{a}+\frac{2}{3}Z_{a}Y^{a}\right. \nonumber\\
&  \left.  +V^{[ab]}\left(  \partial_{a}B_{b}-\partial_{b}B_{a}\right)
-2V^{[ab]}W_{ab}+2W^{bc}\partial_{b}A_{c}-\frac{2}{3}Z_{a}Z^{a}+2Z^{b}%
\partial_{b}S-W^{bc}W_{cb}+\frac{1}{3}W^{2}\right\}  .
\end{align}
Here $W_{\;b}^{b}=W$ and the indices $a,b,c,...$ run from $0$ to $3$. We fix
the gauge parameter $\omega_{ab}$ to eliminate the antisymmetric part of
$e_{bc}$, which yields
\begin{align}
S &  =\frac{1}{2}\int d^{4}x\left\{  Y^{c[ab]}\left(  \partial_{a}e_{\left\{
bc\right\}  }-\partial_{b}e_{\left\{  ac\right\}  }\right)  -Y_{c[ab]}%
Y^{b[ac]}+\frac{1}{3}Y_{a}Y^{a}\right. \nonumber\\
&  .\left.  +V^{[ab]}\left(  \partial_{a}B_{b}-\partial_{b}B_{a}\right)
-2W_{ac}V^{[ac]}-W^{bc}W_{cb}+\frac{1}{3}W_{\;b}^{b}W_{\;c}^{c}+2W^{bc}%
\partial_{b}A_{c}+\frac{2}{3}Z_{a}Y^{a}+2Z^{b}\partial_{b}S-\frac{2}{3}%
Z_{a}Z^{a}\right\}  .
\end{align}
Next we eliminate $Z^{a}$ from the corresponding equation of motion%
\begin{equation}
Z^{a}=\frac{1}{2}\left(  Y^{a}+3\partial^{a}S\right)  ,
\end{equation}
obtaining
\begin{align}
S &  =\frac{1}{2}\int d^{4}x\left\{  Y^{c[ab]}\left(  \partial_{a}e_{\left\{
bc\right\}  }-\partial_{b}e_{\left\{  ac\right\}  }\right)  -Y_{c[ab]}%
Y^{b[ac]}+\frac{1}{2}Y_{a}Y^{a}+Y^{a}\partial_{a}S+\frac{3}{2}\partial
^{a}S\partial_{a}S\right. \nonumber\\
&  \left.  +V^{[ab]}\left(  \partial_{a}B_{b}-\partial_{b}B_{a}\right)
+2W^{bc}\partial_{b}A_{c}-2V^{[ac]}W_{ac}-W^{bc}W_{cb}+\frac{1}{3}%
W^{2}\right\}  .\label{ACTINT}%
\end{align}
Our next step is to redefine
\[
e_{\left\{  bc\right\}  }\rightarrow e_{\left\{  bc\right\}  }-\frac{1}{2}%
\eta_{bc}S,
\]
in such a way to eliminate the crossed term $Y^{a}\partial_{a}S$ so that
(\ref{ACTINT}) reduces to
\begin{align}
S &  =\frac{1}{2}\int d^{4}x\left\{  Y^{c[ab]}\left(  \partial_{a}e_{\left\{
bc\right\}  }-\partial_{b}e_{\left\{  ac\right\}  }\right)  -Y_{c[ab]}%
Y^{b[ac]}+\frac{1}{2}Y_{a}Y^{a}+\frac{3}{2}\partial^{a}S\partial_{a}S\right.
\nonumber\\
&  \left.  +V^{[ab]}\left(  \partial_{a}B_{b}-\partial_{b}B_{a}\right)
-2V^{[ac]}W_{ac}-W^{bc}W_{cb}+\frac{1}{3}W^{2}-2W^{bc}\partial_{b}%
A_{c}\right\}  ,
\end{align}
which already shows the decoupling of the three sectors of the theory. The
above action is invariant under the following gauge transformations, which
basically include only the vector sector%
\begin{align*}
\delta A_{a} &  =\theta_{a},\;\;\delta B_{a}=-\theta_{a},\;\;\delta
V^{[ab]}=0,\\
\delta W_{cb} &  =\partial_{b}\theta_{c}-\eta_{bc}\partial_{a}\theta
^{a},\;\;\;\delta W=-36\ \partial_{a}\theta^{a}\\
\delta e_{\left\{  bc\right\}  } &  =0,\;\delta Y^{[ab]c}=0,\;\delta S=0.
\end{align*}
Next we fix the gauge in two alternative forms that yield the standard
massless action for the spin one field.

\subsection{CASE I}

We choose the parameter $\theta_{a}$ in such a way that $A_{a}=0$, leading to%
\begin{align}
S &  =\frac{1}{2}\int d^{4}x\left\{  Y^{c[ab]}\left(  \partial_{a}e_{\left\{
bc\right\}  }-\partial_{b}e_{\left\{  ac\right\}  }\right)  -Y_{c[ab]}%
Y^{b[ac]}+\frac{1}{2}Y_{a}Y^{a}+\frac{3}{2}\partial^{a}S\partial_{a}S\right.
\nonumber\\
&  \left.  +V^{[ab]}\left(  \partial_{a}B_{b}-\partial_{b}B_{a}-2W_{ab}%
\right)  -W^{bc}W_{cb}+\frac{1}{3}W^{2}\right\}  .\label{ACCASEI}%
\end{align}
Here $V^{[ab]}$ is a Lagrange multiplier, which implies that%
\[
W_{ab}=\frac{1}{2}\left(  \partial_{a}B_{b}-\partial_{b}B_{a}\right)
\rightarrow W=0.
\]
Substituting in (\ref{ACCASEI}) we recover the standard contribution to the
massless spin one field, up to a normalization factor.
\begin{equation}
S=\frac{1}{2}\int d^{4}x\left[  Y^{c[ab]}\left(  \partial_{a}e_{\left\{
bc\right\}  }-\partial_{b}e_{\left\{  ac\right\}  }\right)  -Y_{c[ab]}%
Y^{b[ac]}+\frac{1}{2}Y_{a}Y^{a}+\frac{3}{2}\partial^{a}S\partial_{a}S+\frac
{1}{4}\left[  \partial_{a}B_{b}-\partial_{b}B_{a}\right]  ^{2}\right]  .
\end{equation}

\subsection{CASE\ II}

Now we fix the parameter $\theta_{a}$ such that $;B_{a}=0$ obtaining
\begin{align}
S &  =\frac{1}{2}\int d^{4}x\left\{  Y^{c[ab]}\left(  \partial_{a}e_{\left\{
bc\right\}  }-\partial_{b}e_{\left\{  ac\right\}  }\right)  -Y_{c[ab]}%
Y^{b[ac]}+\frac{1}{2}Y_{a}Y^{a}+\frac{3}{2}\partial^{a}S\partial_{a}S\right.
\nonumber\\
&  \left.  -2V^{[ac]}W_{ac}-W^{bc}W_{cb}+\frac{1}{3}W^{2}-2W^{bc}\partial
_{b}A_{c}\right\}  .
\end{align}
Now the Lagrange multiplier$\ V^{[ac]}$ implies that $W_{ac}=W_{\left\{
ac\right\}  }$ is symmetrical, yielding%
\begin{align}
S &  =\frac{1}{2}\int d^{4}x\left\{  Y^{c[ab]}\left(  \partial_{a}e_{\left\{
bc\right\}  }-\partial_{b}e_{\left\{  ac\right\}  }\right)  -Y_{c[ab]}%
Y^{b[ac]}+\frac{1}{2}Y_{a}Y^{a}+\frac{3}{2}\partial^{a}S\partial_{a}S\right.
\nonumber\\
&  \left.  -W^{\left\{  bc\right\}  }W_{\left\{  cb\right\}  }+\frac{1}%
{3}W^{2}-W^{\left\{  bc\right\}  }\left(  \partial_{b}A_{c}+\partial_{c}%
A_{b}\right)  \right\}  .\label{ACTCASEII}%
\end{align}
Notice that the second line in the above equation must provide an alternative
way of presenting the action for a massless spin one field, even though it is
written in terms of symmetrical fields. We can verify this statement just by
eliminating the field $W^{\left\{  bc\right\}  }$. The corresponding equation
of motion produces%

\begin{equation}
W_{\left\{  bc\right\}  }=-\frac{1}{2}\left(  \partial_{b}A_{c}+\partial
_{c}A_{b}\right)  +\eta_{bc}\partial_{a}A^{a}%
\end{equation}
and the substitution in (\ref{ACTCASEII}) leads indeed to the expected action%
\begin{equation}
S=\frac{1}{2}\int d^{4}x\left\{  Y^{c[ab]}\left(  \partial_{a}e_{\left\{
bc\right\}  }-\partial_{b}e_{\left\{  ac\right\}  }\right)  -Y_{c[ab]}%
Y^{b[ac]}+\frac{1}{2}Y_{a}Y^{a}+\frac{3}{2}\partial^{a}S\partial_{a}S+\frac
{1}{4}\left(  \partial_{b}A_{c}-\partial_{c}A_{b}\right)  ^{2}\right\}
\end{equation}

\begin{acknowledgements}
LFU would like to thanks useful discussions with J. A. Garc\'{\i}a. A. K.
acknowledges institutional support from CDCHT-ULA under project C-1506-07-05-B
and the Program High Energy Physics Latinamerican-European Network (HELEN).
R.M. acknowledges partial support from CONICET-Argentina. L.F.U is partially
supported by projects CONACYT \# 55310 and DGAPA-UNAM-IN109108. R.M. and
L.F.U. have been partially supported by a project of international cooperation
CONACYT-CONICET.
\end{acknowledgements}

\bigskip

\bigskip

\end{document}